\documentclass[aps, prb, preprint, bibnotes, preprintnumbers, amsmath, amssymb, superscriptaddress]{revtex4}

\usepackage{graphicx}
\usepackage{dcolumn}
\usepackage{bm}

\begin{document}

\title{Yttrium Iron Garnet Thin Films with Very Low Damping Obtained by Recrystallization of Amorphous Material}

\author{C. Hauser}
\affiliation{Institute of Physics,
Martin-Luther-Universit\"{a}t Halle-Wittenberg, Von-Danckelmann-Platz 3,
D-06120 Halle, Germany}

\author{T. Richter}
\affiliation{Institute of Physics, Martin-Luther-Universit\"{a}t
Halle-Wittenberg, Von-Danckelmann-Platz 3, D-06120 Halle, Germany}

\author{N. Homonnay}
\affiliation{Institute of Physics, Martin-Luther-Universit\"{a}t
Halle-Wittenberg, Von-Danckelmann-Platz 3, D-06120 Halle, Germany}

\author{C. Eisenschmidt}
\affiliation{Institute of Physics, Martin-Luther-Universit\"{a}t
Halle-Wittenberg, Von-Danckelmann-Platz 3, D-06120 Halle, Germany}

\author{H. Deniz}
\affiliation{Max-Planck-Institut für Mikrostrukturphysik, Weinberg 2, D-06120 Halle, Germany}

\author{D. Hesse}
\affiliation{Max-Planck-Institut für Mikrostrukturphysik, Weinberg 2, D-06120 Halle, Germany}

\author{S. Ebbinghaus}
\affiliation{Institute of Chemistry, Martin-Luther-Universit\"{a}t
Halle-Wittenberg, Kurt-Mothes-Str. 2, D-06120 Halle, Germany}

\author{G. Schmidt}
\email[Correspondence to G. Schmidt: ]{georg.schmidt@physik.uni-halle.de}
\affiliation{Institute of Physics, Martin-Luther-Universit\"{a}t
Halle-Wittenberg, Von-Danckelmann-Platz 3, D-06120 Halle, Germany}
\affiliation{Interdisziplinäres Zentrum für Materialwissenschaften, Martin-Luther University Halle-Wittenberg, Nanotechnikum Weinberg, Heinrich-Damerow-Str. 4, D-06120 Halle, Germany}

\begin{abstract}
We have investigated recrystallization of amorphous Yttrium Iron Garnet (YIG) by annealing in oxygen atmosphere. Our findings show that well below the melting temperature the material transforms into a fully epitaxial layer with exceptional quality, both structural and magnetic.\\
In ferromagnetic resonance (FMR) ultra low damping and extremely narrow linewidth can be observed. For a 56\,nm thick layer a damping constant of $\alpha$\,=\,(6.63\,$\pm$\,1.50)\,$\cdot$\,10$^{-5}$ is found and the linewidth at 9.6\,GHz is as small as 1.30\,$\pm$\,0.05\,Oe which are the lowest values for PLD grown thin films reported so far. Even for a 20\,nm thick layer a damping constant of $\alpha$\,=\,(7.51\,$\pm$\,1.40)\,$\cdot$\,10$^{-5}$ is found which is the lowest value for ultrathin films published so far. The FMR linewidth in this case is 3.49\,$\pm$\,0.10\,Oe at 9.6\,GHz. Our results not only present a method of depositing thin film YIG of unprecedented quality but also open up new options for the fabrication of thin film complex oxides or even other crystalline materials.

\end{abstract}

\flushbottom
\maketitle
\thispagestyle{empty}

\section{Introduction}
YIG can be considered the most prominent material in spin dynamics in thin films and related areas. It is widely used in ferromagnetic resonance experiments\cite{dAllivyKelly.2013, Liu.2014, Sun.2012, Sun.2013, Manuilov.2010, Manuilov.2009}, research on magnonics\cite{Chumak.2014, Kurebayashi.2011, Ustinov.2010, Inoue.2011, Yu.2014, Bi.2014, Demokritov.2003, Chumak.2010} and magnon-based Bose-Einstein-condesates\cite{Li.2013, NowikBoltyk.2012, Demokritov.2006, Serga.2014} because of its exceptionally low damping even in thin films. In research on spin pumping\cite{Ando.2013, Burrowes.2012, Jungfleisch.2013, Rezende.2013, Zutic.2011} and investigation of the inverse spin hall effect\cite{dAllivyKelly.2013, Ando.2013, Sakimura.2014, Castel.2012, Schreier.2015, Kajiwara.2010} it greatly facilitates experiments because it is an insulating material which avoids numerous side effects which occur when ferromagnetic metals are used.\cite{Obstbaum.2014, Mecking.2007} The field of spin caloritronics\cite{Bauer.2012, Sinova.2010, Kirihara.2012, Siegel.2014, Qu.2013, An.2013, Uchida.2010, Adachi.2011, Agrawal.2013} also would not have developed that rapidly without the availability of a non-conducting magnet with long magnon lifetimes.\\
The new fields of applications have resulted in a growing need of high quality thin films, for example for integrated magnonics where layers need to be as thin as 100\,nm or even less. While formerly only micrometer thick films were used which can be obtained by liquid phase epitaxy with very high quality\cite{Gornert.1988, WeiHongxu.1984, Cermak.1990, YongChoi.1998} ultrathin films are nowadays mostly fabricated by pulsed laser deposition (PLD) of epitaxial films at elevated temperature. Especially for ultra thin films (20\,nm or less) grown by PLD quality is high but limited and best results so far show a linewidth in FMR of 2.1\,Oe at 9.6\,GHz.\cite{dAllivyKelly.2013}

\section{Sample Fabrication}
The amorphous YIG layers are deposited on (111) oriented gallium gadolinium garnet (GGG) substrates. After deposition the samples are removed from the PLD chamber and cut into smaller pieces before the subsequent annealing procedure which is done in a quartz oven under pure (99.998\,\%) oxygen atmosphere at ambient pressure at 800$^\circ$C for 30\,minutes (sample\,A, 56\,nm thick), at 800$^\circ$C for three hours (sample\,B, 20\,nm thick), and at 900$^\circ$C for four hours (sample\,C, 113\,nm thick). After annealing the samples are subject to various structural and magnetic characterization experiments.

\section{Structural characterization}
Structural characterization is done by X-ray diffraction, X-ray reflectometry transmission electron microscopy, and Reflection high energy electron diffraction (RHEED).
\subsection{X-ray characterization}
\begin{figure}[H]
\centering
\includegraphics[width=8cm]{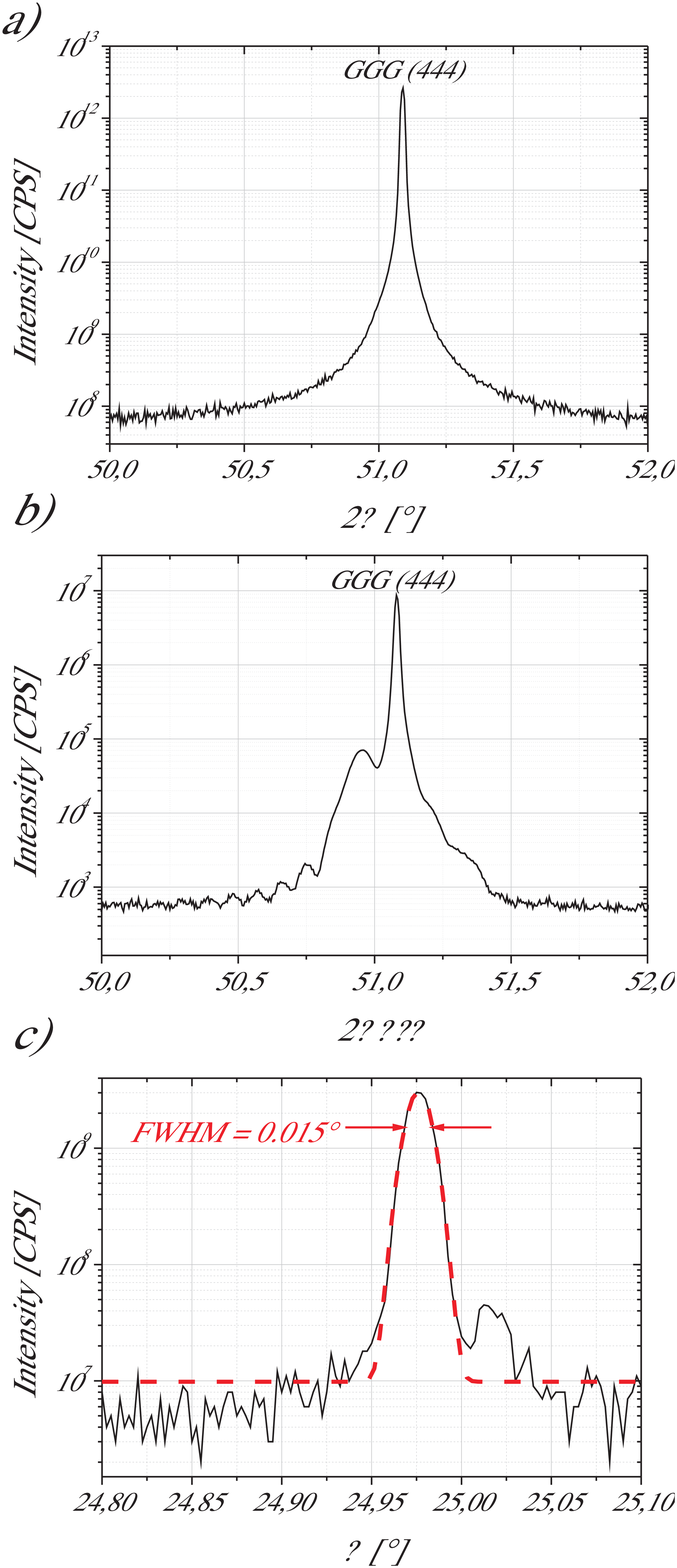}
\caption{X-ray diffraction ($\omega$/2$\theta$ scans) for an unannealed (a) and an annealed YIG layer (b). Before annealing only the substrate peak is visible. After annealing the YIG peak clearly shows up. The position of the peak and the thickness fringes indicate fully pseudomorphic growth and smooth interfaces. (c) shows a rocking curve of the layer peak shown in (b). The full width at half maximum is only 0.015$^\circ$. The dotted line shows a Gaussian fit to the peak.}\label{XRD}
\end{figure}

X-ray diffraction is performed by doing an $\omega$/2$\theta$ scan of the (444) reflex and a rocking curve of the YIG layer peak. Before annealing the diffraction pattern (Figure\,\ref{XRD}a) only shows the peak of the GGG substrate indicating an amorphous or at least highly polycrystalline YIG film. A truly amorphous nature is confirmed by transmission electron microscopy as described below. After annealing, the diffraction pattern is completely changed. Figure\,\ref{XRD}b shows the $\omega$/2$\theta$ scan for sample\,C. Here we clearly observe the diffraction peak of the YIG film at an angle corresponding to the small lattice mismatch of YIG on GGG which is only 0.057\,\%. Even thickness fringes can be observed indicating a very smooth layer with low interface and surface roughness. The layer peak is further investigated in a rocking curve (Figure\,\ref{XRD}c) which shows a full width at half maximum (FWHM) of 0.015$^\circ$ indicating a fully pseudomorphic YIG layer. Roughness is also crosschecked using X-ray reflectometry showing an RMS value of less than 0.2\,nm\cite{Lang.2014}. It should, however, be noted that for not-annealed layers the RMS roughness is even smaller than 0.1\,nm.

\subsection{Transmission electron microscopy}
\begin{figure}[H]
\centering
\includegraphics[width=6.8cm]{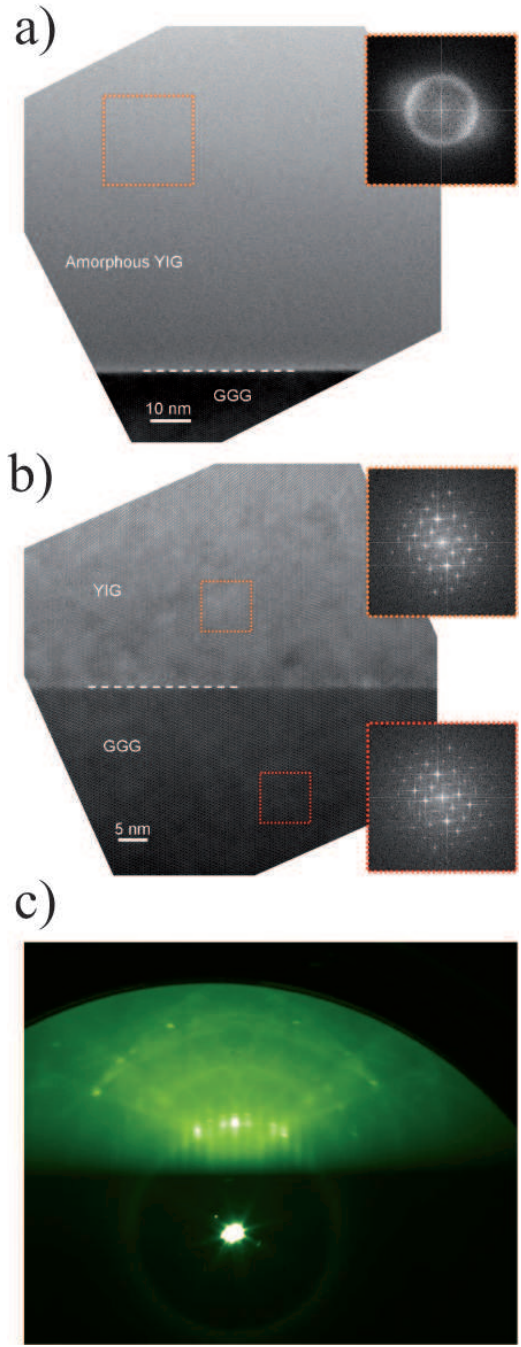}
\caption{a) A high resolution TEM (HRTEM) image of an amorphous YIG film on GGG substrate. The inset shows a FFT pattern from the region of interest (dotted frame) in the amorphous layer. b) A HRTEM image of the interface between the annealed YIG film and the GGG substrate of sample\,C. The insets show FFT patterns from the regions of interest in the film and the substrate. The YIG film exhibits epitaxial growth with respect to the substrate and appears monocrystalline. c) RHEED image obtained from the surface of an annealed YIG film (sample\,B). Kikuchi lines\cite{Braun.1999} indicate a two dimensional highly ordered surface.
}\label{TEM}
\end{figure}

For Transmission electron microscopy (TEM) preparation the sample surface is protected by depositing a thin Pt layer. Then thin lamellae are cut out using focused ion beam preparation. The orientation of the samples is chosen for cross sectional TEM along the cubic crystalline axis. TEM is performed using a JEOL JEM-4010 electron microscope at an acceleration voltage of 400\,kV. For the nominally amorphous sample the pictures (Figure\,\ref{TEM}a) show a pure film without inclusions but also without any trace of polycristallinity. Further analysis using fast fourier transform confirms that the YIG layer is indeed completely amorphous. For an annealed sample (sample\,C) the result of the TEM investigation is surprising (Figure\,\ref{TEM}b). The sample is not only monocrystalline but it also shows no sign of inclusions or defects and even the interface to the GGG appears flawless.

\subsection{Reflection high energy electron diffraction}
The atomic order of the layer surface after annealing is further investigated by Reflection high energy electron diffraction (RHEED). For this purpose sample\,B is again introduced into the PLD chamber after the annealing process. After evacuation a clear RHEED pattern is observed. The RHEED image (Figure\,\ref{TEM}c) not only shows the typical pattern for a YIG surface during high temperature growth but also exhibits the so called Kikuchi lines.\cite{Braun.1999} We do not observe these lines in high temperature growth of epitaxial YIG. They are typically a sign of a surface of excellent two dimensional growth, again indicating that the crystalline quality of the annealed layers is extremely high.

\section{Magnetic characterization}
Magnetic characterization is done using SQUID magnetometry and FMR at room temperature.
\subsection{SQUID magnetometry}
\begin{figure}[H]
\centering
\includegraphics[width=10cm]{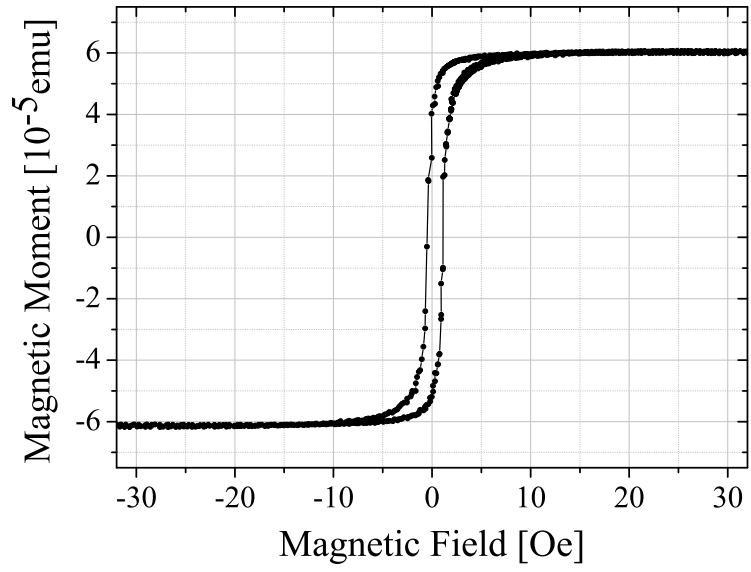}
\caption{Hysteresis loop as measured by SQUID magnetometry for a 113\,nm thick YIG sample after annealing. The paramagnetic background caused by the GGG substrate was subtracted.}\label{SQUID}
\end{figure}

In SQUID magnetometry hysteresis loops are taken on sample\,C. The data is corrected by subtracting a linear paramagnetic contribution which is caused by the GGG substrate. After correction the observed saturation magnetization is (105\,$\pm$\,3)\,emu\,cm$^{-3}$ which is approx. 30\,\% below the bulk value\cite{Hansen.1974} (Figure\,\ref{SQUID}). The coercive field is determined as (0.8\,$\pm$\,0.1)\,Oe.

\subsection{Ferromagnetic resonance}
\begin{figure}[H]
\centering
\includegraphics[width=16cm]{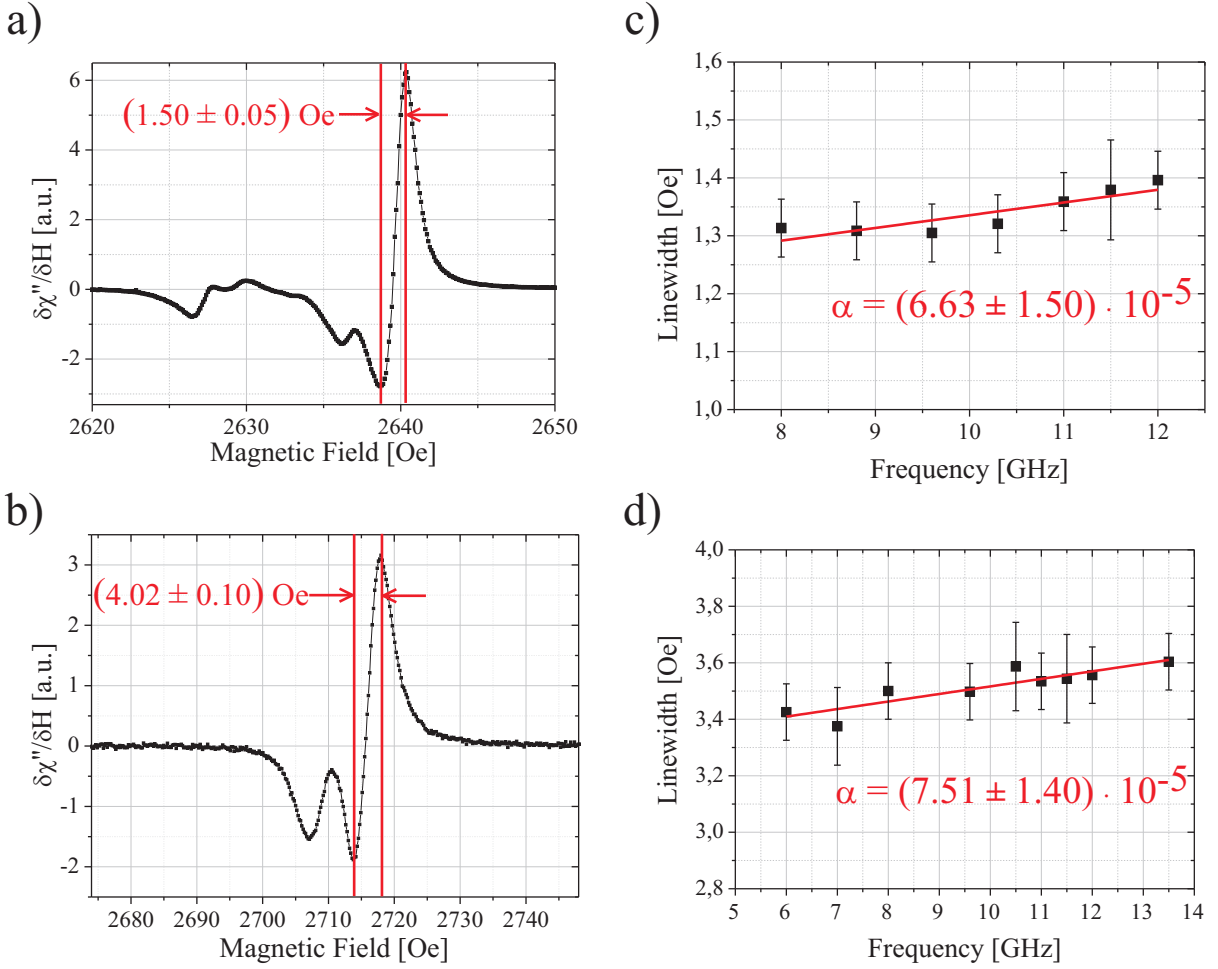}
\caption{a) and b) FMR data obtained at 9.6\,GHz for a 56\,nm thick (a, sample\,A) and a 20\,nm thick (b, sample\,B) YIG layer after annealing. The main resonance lines have a peak-to-peak linewidth of 1.51\,$\pm$\,0.05 Oe (sample\,A) and 4.02\,$\pm$\,0.10 Oe (sample\,B). This peak-to-peak linewidth corresponds to a true linewidth of 1.30\,$\pm$\,0.05 Oe and 3.49\,$\pm$\,0.10 Oe, respectively.
c) and d) Frequency dependence of the FMR linewidth for sample\,A and sample\,B. The fits are a straight line corresponding to a damping of $\alpha$\,=\,(6.63\,$\pm$\,1.50)\,$\cdot$\,10$^{-5}$ (c, sample\,A) and $\alpha$\,=\,(7.51\,$\pm$\,1.40)\,$\cdot$\,10$^{-5}$ (d, sample\,B).}\label{FMR}
\end{figure}

FMR is performed by putting the samples face down on a coplanar waveguide whose magnetic radio frequency (RF) field is used for excitation. The setup is placed in a homogenous external magnetic field which is superimposed with a small low frequency modulation. RF absorption is measured using a lock-in amplifier. As expected no signal can be detected for unannealed YIG layers. For annealed samples a clear resonance is observed. Figure\,\ref{FMR}a shows the resonance signal for sample\,A. The linewidth which is obtained by multiplying the peak to peak linewidth of the derivative of the absorption by a factor of $\sqrt{3}/2$\cite{Sun.2012, Liu.2014, HouchenChang.2014, GeorgWoltersdorf.August2004} is only 1.30\,$\pm$\,0.05 Oe at 9.6\,GHz which is the smallest value for thin films reported so far.\cite{dAllivyKelly.2013} In Figure\,\ref{FMR}b the resonance of sample\,B is shown. Here the linewidth at 9.6\,GHz is 3.49\,$\pm$\,0.10\,Oe. For sample\,C the linewidth is 1.65\,$\pm$\,0.10\,Oe at 9.6\,GHz (no figure).\\
In order to determine the damping constant $\alpha$ frequency dependent measurements are performed on sample\,A. The excitation frequency is varied between 8 and 12\,GHz. Results are plotted in Figure\,\ref{FMR}c. As described by Chang \textit{et al.}\cite{HouchenChang.2014} and Liu \textit{et al.}\cite{Liu.2014} we first determine the gyromagnetic ratio of $\gamma$\,=\,(2.92\,$\pm$\,0.01)\,MHz\,Oe$^{-1}$ and a linewidth at zero magnetic field of approx. 1.11\,$\pm$\,0.05 Oe. The damping can then be calculated from the frequency dependence of the linewidth to $\alpha$\,=\,(6.63\,$\pm$\,1.50)\,$\cdot$\,10$^{-5}$. This damping is even lower than the lowest value reported by Chang et al.\cite{HouchenChang.2014}. It is interesting to note that Chang \textit{et al.} did not observe a similarly small linewidth for their layer.\cite{HouchenChang.2014} d’Allivy Kelly \textit{et al.} on the other hand do observe a smaller linewidth for 20\,nm thick layers of 2.1\,Oe at 9.6\,GHz\cite{dAllivyKelly.2013}, however, the damping they find is three times as big as in our case. For sample\,B (20\,nm) we found a gyromagnetic ratio of $\gamma$\,=\,(2.81\,$\pm$\,0.01)\,MHz\,Oe$^{-1}$ and a linewidth at zero magnetic field of approx. 3.25\,$\pm$\,0.05 Oe and the damping was determined as $\alpha$\,=\,(7.51\,$\pm$\,1.40)\,$\cdot$\,10$^{-5}$ which is also lower than any other value reported for similarly thin films (Figure\,\ref{FMR}d).

\section{Discussion}
In conclusion we can state that using high temperature annealing in oxygen atmosphere it is possible to transform amorphous YIG layers of tens of nanometers of thickness into epitaxial thin films with extremely small FMR linewidth and exceptionally low damping. The crystalline quality is extremely high. Our findings may thus present a new and easy route for thin film fabrication of epitaxial complex oxides.

\section{Methods}
The substrates are prepared by cleaning in acetone and isopropanol and then introduced into the PLD chamber which is copper sealed and UHV compatible with a background pressure as low as 10$^{-9}$ mbar. Deposition is done at an oxygen partial pressure of 0.025\,mbar with the substrate at room temperature. A laser with a typical fluency of 2.5\,J\,cm$^{-2}$ and a wavelength of 248\,nm is used at a repetition rate of 5\,Hz. The target is a stoichiometric YIG target prepared in-house. With these parameters we obtain a growth rate of approx. 0.5\,nm\,min$^{-1}$.

\section*{Acknowledgements}
This work was supported by the European Commission in the project IFOX under grant agreement NMP3-LA-2010-246102 and by the DFG in the SFB\,762. We thank Georg Woltersdorf and Sergey Manuilow for fruitful discussion.


\end{document}